\documentclass[conference]{IEEEtran}

\IEEEoverridecommandlockouts
\usepackage{cite}
\usepackage{amsmath,amssymb,amsfonts}
\usepackage{algorithm,algorithmic}
\usepackage{graphicx}
\usepackage{textcomp}
\usepackage{xcolor}
\usepackage[acronym]{glossaries}
\usepackage{blindtext}
\def\BibTeX{{\rm B\kern-.05em{\sc i\kern-.025em b}\kern-.08em
    T\kern-.1667em\lower.7ex\hbox{E}\kern-.125emX}}
\usepackage{booktabs}
\usepackage{bm}
\usepackage{import}
\usepackage{physics}
\usepackage[hidelinks]{hyperref}
\usepackage{url}
\usepackage{multirow}
\usepackage{blindtext}
\usepackage{makecell}
\usepackage{pythonhighlight}
\usepackage{nicefrac}
\usepackage{caption}
\usepackage{lipsum}
\usepackage{multirow}
\usepackage{tabularx}
\usepackage{afterpage}
\usepackage{orcidlink}

\usepackage{pgfplots}
\pgfplotsset{compat=1.14}

\usepackage[raggedright]{subfigure}

\usepackage[capitalize,noabbrev]{cleveref}
\Crefname{equation}{Eq.}{Eqs.}
\Crefname{figure}{Fig.}{Figs.}
\Crefname{section}{Sec.}{Secs.}
\Crefname{table}{Tab.}{Tabs.}

\newcommand\copyrighttext{%
  \footnotesize \textcopyright 2026 IEEE. Personal use of this material is permitted.
  Permission from IEEE must be obtained for all other uses, in any current or future
  media, including reprinting/republishing this material for advertising or promotional
  purposes, creating new collective works, for resale or redistribution to servers or
  lists, or reuse of any copyrighted component of this work in other works.}
\newcommand\copyrightnotice{%
\begin{tikzpicture}[remember picture,overlay]
\node[anchor=south,yshift=10pt] at (current page.south) {\fbox{\parbox{\dimexpr\textwidth-\fboxsep-\fboxrule\relax}{\copyrighttext}}};
\end{tikzpicture}%
}

\begin{document}

\newacronym{vqc}{VQC}{variational quantum circuit}
\newacronym{cptp}{CPTP}{completely positive trace preserving}
\newacronym{qec}{QEC}{quantum error correction}
\newacronym{varqec}{VarQEC}{variational quantum error correction}
\newacronym{rea}{REA}{randomized entangling ansatz}
\newacronym{ml}{ML}{machine learning}

\title{Learning to Concatenate Quantum Codes
    \thanks{The research was supported by the German Federal Ministry of Research, Technology and Space, funding program Quantum Systems, via the project Q-GeneSys, grant number 13N17389. The research is also part of the Munich Quantum Valley (MQV), which is supported by the Bavarian state government with funds from the Hightech Agenda Bayern Plus.\\
    Correspondence to: nico.meyer@iis.fraunhofer.de
    }
}

\author{
    \IEEEauthorblockN{Nico Meyer\IEEEauthorrefmark{1}\IEEEauthorrefmark{2}\orcidlink{0000-0002-5463-5437}, Christopher Mutschler\IEEEauthorrefmark{1}\IEEEauthorrefmark{3}\orcidlink{0000-0001-8108-0230}, Dominik Seu{\ss}\IEEEauthorrefmark{1}\IEEEauthorrefmark{4}\orcidlink{0000-0001-5302-2236}, Andreas Maier\IEEEauthorrefmark{2}\orcidlink{0000-0002-9550-5284}, and Daniel D.\ Scherer\IEEEauthorrefmark{1}\orcidlink{0000-0003-0355-4140}
    }
    \IEEEauthorblockA{
        \IEEEauthorrefmark{1}Fraunhofer IIS, Fraunhofer Institute for Integrated Circuits IIS, Nuremberg, Germany\\
        \IEEEauthorrefmark{2}Pattern Recognition Lab, Friedrich-Alexander-University Erlangen-Nuremberg, Erlangen, Germany\\
        \IEEEauthorrefmark{3}Machine Learning and Positioning Systems Lab, University of Technology Nuremberg (UTN), Nuremberg, Germany\\
        \IEEEauthorrefmark{4}Center for Artificial Intelligence (CAIRO), Technical University of Applied Sciences W\"urzburg-Schweinfurt, Germany
    }
}

\maketitle

\begin{abstract}
    Concatenating quantum error correction codes scales error correction capability by driving logical error rates down double-exponentially across levels. However, the noise structure shifts under concatenation, making it hard to choose an optimal code sequence. We automate this choice by estimating the effective noise channel after each level and selecting the next code accordingly. In particular, we use learning-based methods to tailor small, non-additive encoders when the noise exhibits sufficient structure, then switch to standard codes once the noise is nearly uniform. In simulations, this level-wise adaptation achieves a target logical error rate with far fewer qubits than concatenating stabilizer codes alone--reducing qubit counts by up to two orders of magnitude for strongly structured noise. Therefore, this hybrid, learning-based strategy offers a promising tool for early fault-tolerant quantum computing.
\end{abstract}

\begin{IEEEkeywords}
quantum computing, machine learning, quantum error correction, code concatenation, variational algorithm
\end{IEEEkeywords}

\copyrightnotice  
\vspace{-0.5cm} 

\glsresetall
\section{\label{sec:intro}Introduction}

\Gls{qec} is essential for reliable quantum computation in the presence of noise and decoherence. Code concatenation by recursively encoding logical states using nested codes achieves doubly exponential suppression of errors~\cite{knill1998resilient,aliferis2005quantum}, assuming beyond-threshold physical components. Using heterogeneous codes across levels can reduce qubit overhead~\cite{yoshida2025concatenate}, assuming simple stationary noise structure. Realistic noise typically is structured, e.g., dephasing-dominant or otherwise anisotropic~\cite{erhard2019characterizing,krantz2019quantum}. Furthermore, the effective noise channel is reshaped across concatenation levels~\cite{knill1998resilient,aliferis2005quantum,huang2019robustness}, motivating a level-wise, noise-tailored concatenation strategy.

\Gls{ml} can be utilized to discover new \gls{qec} codes with specific properties, in particular using techniques from reinforcement learning~\cite{fosel2018reinforcement,nautrup2019optimizing} and variational methods~\cite{johnson2017qvector,locher2023quantum}. In particular, there also exist \gls{ml} techniques for tailoring encodings to noise structures~\cite{olle2024simultaneous,olle2025scaling,meyer2025learning}. One approach to \gls{varqec}~\cite{cao2022quantum} learns non-additive, measurement-free codes~\cite{heussen2024measurement} by minimizing the information loss under given noise structures, enabling approximate \gls{qec}~\cite{schumacher2002approximate} with fewer qubits than standard stabilizer codes~\cite{meyer2025learning,meyer2025variational}. Yet, how to scale these codes by leveraging the tailored encoders systematically within concatenation is an open research question.

\begin{figure}[tb]
    \centering
    \input{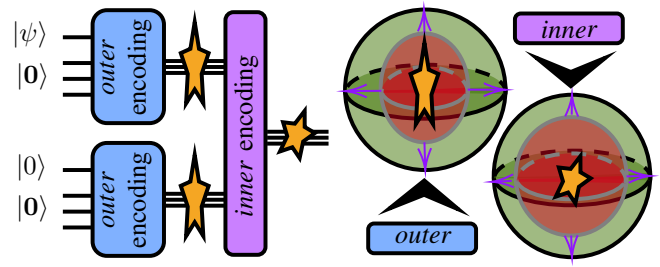}
    \caption{\label{fig:figure1}
    Schematic of noise-aware code concatenation. A logical state $\ket{\psi}$ is recursively encoded by an outer and inner code. Because the effective noise channel changes across levels, we estimate its structure after each concatenation step and use machine-learning methods to tailor the next-level code. This enhances per-level noise suppression and reduces the qubit overhead required to reach a target logical error rate.}
\end{figure}

We address this gap by introducing \emph{learning to concatenate}: a pipeline that alternates between (i) estimating the effective single-qubit logical channel produced by the current level and (ii) tailoring the next-level code to the uncovered noise structures. This procedure is sketched in \cref{fig:figure1}. For the first step, we develop a fidelity-only, two-design estimator to recover the Pauli-Liouville diagonal of the logical channel without full process tomography~\cite{ambainis2007quantum,watrous2018theory,greenbaum2015introduction}. We then tailor small \gls{varqec} patches~\cite{meyer2025learning} when structure is exploitable and switch to standard fixed-distance stabilizer codes once the logical channel is effectively depolarizing. Empirical analyses of this procedure confirms prior findings that anisotropic noise converges rapidly to an isotropic channel under concatenation with non-CSS codes~\cite{huang2019robustness,laflamme1996perfect}. Moreover, we identify noise structures that yield a $4$-fold to over $100$-fold reduction in overhead for achieving a target error rate, compared to concatenating standard codes.

The remainder of this paper is organized as follows: In \cref{sec:preliminaries}, we summarize the necessary background and prior work on quantum error correction, variational codes, and code concatenation. The procedure for estimating the structure of the effective noise channels between concatenation levels, and the learning and concatenation procedure for the tailored codes is outlined in \cref{sec:method}. The empirical analysis of the proposed pipeline for different noise structures is to be found in \cref{sec:empirical}. Finally, in \cref{sec:discussion}, we provide a summary, discuss open questions and future work, and position the proposed concept within the context of early fault-tolerant quantum computing~\cite{katabarwa2024early}.


\section{\label{sec:preliminaries}Preliminaries and Prior Work}

Noise in open quantum systems can be described using the notion of density matrices $\rho$, which describe the \emph{mixed states} as a statistical ensemble of \emph{pure} $n$-qubit states.
The non-unitary process of noise is described by \gls{cptp} maps, typically specified using the \emph{Kraus representation} $\mathcal{N}(\rho) = \sum_{k} E_k \rho E_{k}^{\dagger}$, where the $E_k$ are Kraus operators with $\sum_k E_{k}^{\dagger} E_k = I$~\cite{kraus1971general}.

The most common example of such a noise channel is \emph{symmetric depolarizing noise}, where all error types (i.e.,\ bitflip, phaseflip, and combinations thereof) are equally likely, described in its single-qubit version by
\begin{align}
    \label{eq:noise_depolarizing}
    \mathcal{N}_{\mathrm{dep}}(\rho) = (1-p) \rho + \frac{p}{3} X \rho X^{\dagger} + \frac{p}{3} Y \rho Y^{\dagger} + \frac{p}{3} Z \rho Z^{\dagger},
\end{align}
where $p$ is the overall depolarizing probability. In this work, we target more general noise channels, subsumed under arbitrary single-qubit \emph{Pauli noise}
\begin{align}
    \label{eq:pauli_noise}
    \mathcal{N}_{\mathrm{Pauli}}(\rho) = (1-p) \rho + p_X X \rho X^{\dagger} + p_Y Y \rho Y^{\dagger} + p_Z Z \rho Z^{\dagger},
\end{align}
where the overall noise strength is given by $p = p_X + p_Y + p_Z$. In particular, we focus on three noise channels: (1) \emph{asymmetric depolarizing} noise $\mathcal{N}_{\mathrm{adep}}$ as investigated in~\cite{olle2024simultaneous,meyer2025learning}, with an asymmetry factor $c=0.5$, resulting in $\frac{p_X}{p}=\frac{p_Y}{p}=0.07$ and $\frac{p_Z}{p}=0.86$; (2) standard \emph{bitflip} noise $\mathcal{N}_{\mathrm{bit}}$, with $p_X = p$ and $p_Y = p_Z = 0$; (3) strictly correlated Pauli $X$ and $Z$ noise, i.e.\ a Pauli \emph{$Y$-flip} $\mathcal{N}_{\mathrm{yflip}}$, with $p_Y = p$ and $p_X = p_Z = 0$. While addressing non-Pauli channels like \emph{amplitude damping} noise is conceptually possible~\cite{meyer2025learning}, we leave such considerations for future work.
Throughout this manuscript, we assume that errors for multi-qubit systems act independently on each qubit, giving an overall noise model for an $n$-qubit system as $\mathcal{N}^{\otimes n} = \bigotimes\nolimits_{j=1}^{n} \mathcal{N}_j$. Related literature suggests that our analysis should extend to correlated errors~\cite{meyer2025learning}, which is, however, out of scope for this work.

\subsection{\label{subsec:qec}Quantum Error Correction}

\Glsentrylong{qec} is the primarily pursued concept to protect quantum states against noise channels, by encoding the \emph{logical} state into a collection of physical qubits~\cite{shor1995scheme,steane1996error}. Typically, we refer to \gls{qec} codes $\mathcal{C}$ by their parameters
\begin{align}
    ((n,K,d)).
\end{align}
Hereby, $n$ indicates the number of physical qubits the logical $K$-dimensional state is encoded into, with particular relevance of codes where $k=\log_2K$ is a whole number of logical qubits. The code distance $d$ indicates that up to $d-1$ arbitrary errors can be detected, and $\lfloor \frac{d-1}{2} \rfloor$ arbitrary errors can be corrected. When the code does not have a provable code distance, or it is unknown, we simply write $((n,K))$.
For the special and frequently considered case of stabilizer (i.e. \emph{additive}) codes~\cite{gottesman1997stabilizer}, one typically switches to the notation
\begin{align}
    [[n,k,d]].
\end{align}
Again, $n$ indicates the number of physical qubits, $k$ the number of logical qubits with $K=2^k$, and $d$ the code distance.

The procedure of \gls{qec} can be roughly separated into three consecutive steps, with the first one being the \emph{encoding} of the logical state into multiple physical qubits as
\begin{align}
    \label{eq:encoding}
    \rho_L = U_{\mathrm{enc}} \left( \rho \otimes \ket{0} \bra{0}^{\otimes n-k} \right) U_{\mathrm{enc}}^{\dagger}.
\end{align}
The noise-affected logical state $\tilde{\rho}_L = \mathcal{N}(\rho_L)$ then undergoes potentially multiple rounds of \emph{recovery}
\begin{align}
    \label{eq:recovery}
    \hat{\rho}_L = \mathrm{Tr}_{r} \left( U_{\text{rec}} \left( \tilde{\rho}_L \otimes \ket{0} \bra{0}^{\otimes r} \right) U_{\text{rec}}^{\dagger} \right),
\end{align}
where $r$ is a fresh register of ancilla qubits for every recovery cycle. As we elaborate in \cref{subsec:varqec}, the \gls{qec} procedure in this work is measurement-free~\cite{heussen2024measurement}, i.e.,\ avoids the typical stabilizer measurements and conditional correction operations~\cite{steane1996error}.
Finally, before measurement, the logical state is \emph{decoded} as
\begin{align}
    \label{eq:decoding}
    \hat{\rho} = \mathrm{Tr}_{n-k} \left( U_{\text{enc}}^{\dagger} \hat{\rho}_L U_{\text{enc}} \right).
\end{align}
This gives rise to the overall error correction procedure $\mathcal{R}$, composed from encoding following \cref{eq:encoding}, potentially multiple rounds of recovery according to \cref{eq:recovery}, and final decoding as in \cref{eq:decoding}. The objective of this procedure, associated with an \gls{qec} code $\mathcal{C}$, is to keep the effect of noise under control, i.e.\ $(\mathcal{R} \circ \mathcal{N}) (\rho) \propto \rho$.

\subsection{\label{subsec:varqec}Variational Quantum Error Correction}

Recent work on \gls{ml}-enhanced \gls{qec} has shown that the fit of a \gls{qec} encoding for a specific noise channel can be quantified by the \emph{worst-case distinguishability loss}
\begin{align}
    \label{eq:dloss_worst}
    \overline{\mathcal{D}}(\mathcal{N}) = \max_{\rho,\sigma} \Delta_{T} (\rho,\sigma;\mathcal{N}),
\end{align}
where $\Delta_{T}(\rho,\sigma;\mathcal{N}) = T(\rho,\sigma) - T(\mathcal{N}(\rho_L),\mathcal{N}(\sigma_L))$ is the \emph{lost trace distance} for state pairs $\rho,\sigma$~\cite{meyer2025learning,meyer2025variational}. The intuition is that this measure, based on the trace distance, quantifies the loss of information under encoding and error channel, which should be kept as minimal as possible to allow for successful error correction. This connection has been formally and empirically supported by showing that a low distinguishability loss guarantees the existence of a high-fidelity recovery operation. Furthermore, the quality of a sophisticated recovery operation can be quantified by the \emph{worst-case fidelity loss}
\begin{align}
    \label{eq:floss_worst}
    \overline{\mathcal{F}}(\mathcal{N}) = 1 - \min_{\rho} F(\rho,\hat{\rho}),
\end{align}
with $\hat{\rho}$ as defined in \cref{eq:encoding,eq:recovery,eq:decoding}~\cite{johnson2017qvector}.

To discover encodings $U_{\mathrm{enc}}$ and recovery operations $U_{\mathrm{rec}}$ that are desirable following \cref{eq:dloss_worst,eq:floss_worst}, one can establish a machine learning procedure~\cite{meyer2025learning}: In order to make the operations tunable, they are instantiated as variational quantum circuits~\cite{cerezo2021variational} with trainable parameters $\Theta$ and $\Phi$, respectively. Concretely, we employ a so-called \gls{rea}~\cite{meyer2025learning} for both $U_{\mathrm{enc}}(\Theta)$ and $U_{\mathrm{rec}}(\Phi)$ throughout this work, which consists of an initial parameterized single-qubit layer, followed by randomly-placed parameterized two-qubit operations. Due to this instantiation of encoding and recovery with parameterized circuits, the procedure is also referred to as \glsentrylong{varqec} (VarQEC). 
We note that the acronym ``VarQEC'' was originally introduced for a variational procedure whose objective is derived from the Knill-Laflamme conditions~\cite{cao2022quantum}. However, this is not the notion we employ in this work.

Given a noise channel $\mathcal{N}$, the encoding can be tailored towards the structure of the noise by updating towards
\begin{align}
    \min_{\Theta} \overline{\mathcal{D}}(\mathcal{N};\Theta).
\end{align}
After the encoding has been established, we can proceed to train the measurement-free~\cite{heussen2024measurement} recovery operation by
\begin{align}
    \label{eq:fidelity_loss_simplified}
    \min_{\Phi} \overline{\mathcal{F}}(\mathcal{N};\Phi),
\end{align}
where we restrict to single rounds of recovery. In practice, due to instabilities caused by the extrema operations within the loss function~\cite{hastie2009elements}, the worst-case formulation is replaced by an average-case proxy. Furthermore, to make the evaluation more efficient, a two-design approximation~\cite{ambainis2007quantum,dankert2009exact} is employed. This procedure can be used to discover non-additive \gls{qec} codes that reduce the loss of information under structured noise using less physical qubits than established stabilizer codes~\cite{meyer2025learning}. To further scale the correction capabilities of these \gls{varqec} codes, we envision code concatenation to be an important cornerstone.

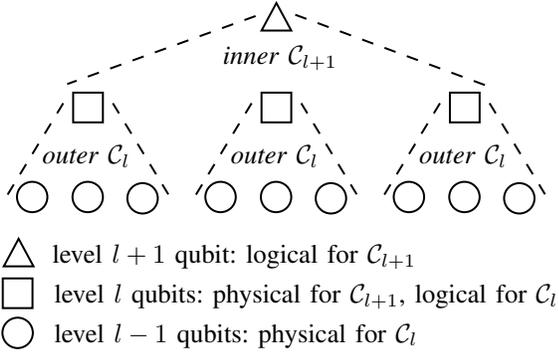
\begin{figure}[tb]
    \centering
    \tikzset{every picture/.style={line width=0.75pt}} 

\begin{tikzpicture}[x=0.75pt,y=0.75pt,yscale=-1,xscale=1]

\draw   (44.67,127.67) .. controls (44.67,123.43) and (48.1,120) .. (52.33,120) .. controls (56.57,120) and (60,123.43) .. (60,127.67) .. controls (60,131.9) and (56.57,135.33) .. (52.33,135.33) .. controls (48.1,135.33) and (44.67,131.9) .. (44.67,127.67) -- cycle ;
\draw   (73,75) -- (88,75) -- (88,90) -- (73,90) -- cycle ;
\draw   (175.5,29.75) -- (183,43.67) -- (168,43.67) -- cycle ;
\draw  [dash pattern={on 4.5pt off 4.5pt}]  (68,80.33) -- (38,129.67) ;
\draw   (72.67,127.67) .. controls (72.67,123.43) and (76.1,120) .. (80.33,120) .. controls (84.57,120) and (88,123.43) .. (88,127.67) .. controls (88,131.9) and (84.57,135.33) .. (80.33,135.33) .. controls (76.1,135.33) and (72.67,131.9) .. (72.67,127.67) -- cycle ;
\draw   (100.33,128.33) .. controls (100.33,124.1) and (103.77,120.67) .. (108,120.67) .. controls (112.23,120.67) and (115.67,124.1) .. (115.67,128.33) .. controls (115.67,132.57) and (112.23,136) .. (108,136) .. controls (103.77,136) and (100.33,132.57) .. (100.33,128.33) -- cycle ;
\draw  [dash pattern={on 4.5pt off 4.5pt}]  (93,80.33) -- (121,126.33) ;
\draw   (139.67,127.67) .. controls (139.67,123.43) and (143.1,120) .. (147.33,120) .. controls (151.57,120) and (155,123.43) .. (155,127.67) .. controls (155,131.9) and (151.57,135.33) .. (147.33,135.33) .. controls (143.1,135.33) and (139.67,131.9) .. (139.67,127.67) -- cycle ;
\draw   (168,75) -- (183,75) -- (183,90) -- (168,90) -- cycle ;
\draw  [dash pattern={on 4.5pt off 4.5pt}]  (163,80.33) -- (133,129.67) ;
\draw   (167.67,127.67) .. controls (167.67,123.43) and (171.1,120) .. (175.33,120) .. controls (179.57,120) and (183,123.43) .. (183,127.67) .. controls (183,131.9) and (179.57,135.33) .. (175.33,135.33) .. controls (171.1,135.33) and (167.67,131.9) .. (167.67,127.67) -- cycle ;
\draw   (195.33,128.33) .. controls (195.33,124.1) and (198.77,120.67) .. (203,120.67) .. controls (207.23,120.67) and (210.67,124.1) .. (210.67,128.33) .. controls (210.67,132.57) and (207.23,136) .. (203,136) .. controls (198.77,136) and (195.33,132.57) .. (195.33,128.33) -- cycle ;
\draw  [dash pattern={on 4.5pt off 4.5pt}]  (188,80.33) -- (216,126.33) ;
\draw   (234.67,127.67) .. controls (234.67,123.43) and (238.1,120) .. (242.33,120) .. controls (246.57,120) and (250,123.43) .. (250,127.67) .. controls (250,131.9) and (246.57,135.33) .. (242.33,135.33) .. controls (238.1,135.33) and (234.67,131.9) .. (234.67,127.67) -- cycle ;
\draw   (263,75) -- (278,75) -- (278,90) -- (263,90) -- cycle ;
\draw  [dash pattern={on 4.5pt off 4.5pt}]  (258,80.33) -- (228,129.67) ;
\draw   (262.67,127.67) .. controls (262.67,123.43) and (266.1,120) .. (270.33,120) .. controls (274.57,120) and (278,123.43) .. (278,127.67) .. controls (278,131.9) and (274.57,135.33) .. (270.33,135.33) .. controls (266.1,135.33) and (262.67,131.9) .. (262.67,127.67) -- cycle ;
\draw   (290.33,128.33) .. controls (290.33,124.1) and (293.77,120.67) .. (298,120.67) .. controls (302.23,120.67) and (305.67,124.1) .. (305.67,128.33) .. controls (305.67,132.57) and (302.23,136) .. (298,136) .. controls (293.77,136) and (290.33,132.57) .. (290.33,128.33) -- cycle ;
\draw  [dash pattern={on 4.5pt off 4.5pt}]  (283,80.33) -- (311,126.33) ;
\draw  [dash pattern={on 4.5pt off 4.5pt}]  (165,36.33) -- (70,70.67) ;
\draw  [dash pattern={on 4.5pt off 4.5pt}]  (185,35.67) -- (282,71.67) ;
\draw   (45.5,148.75) -- (53,162.67) -- (38,162.67) -- cycle ;
\draw   (37.5,168.75) -- (52.5,168.75) -- (52.5,183.75) -- (37.5,183.75) -- cycle ;
\draw   (37.33,196.25) .. controls (37.33,192.02) and (40.77,188.58) .. (45,188.58) .. controls (49.23,188.58) and (52.67,192.02) .. (52.67,196.25) .. controls (52.67,200.48) and (49.23,203.92) .. (45,203.92) .. controls (40.77,203.92) and (37.33,200.48) .. (37.33,196.25) -- cycle ;

\draw (56,100) node [anchor=north west][inner sep=0.75pt]   [align=left] {\textit{outer} $\displaystyle \mathcal{C}_{l}$};
\draw (151,100) node [anchor=north west][inner sep=0.75pt]   [align=left] {\textit{outer} $\displaystyle \mathcal{C}_{l}$};
\draw (246,100) node [anchor=north west][inner sep=0.75pt]   [align=left] {\textit{outer} $\displaystyle \mathcal{C}_{l}$};
\draw (147,50) node [anchor=north west][inner sep=0.75pt]   [align=left] {\textit{inner} $\displaystyle \mathcal{C}_{l+1}$};
\draw (61,150) node [anchor=north west][inner sep=0.75pt]   [align=left] {level $\displaystyle l+1$ qubit: logical for $\displaystyle \mathcal{C}_{l+1}$ };
\draw (62,170) node [anchor=north west][inner sep=0.75pt]   [align=left] {level $\displaystyle l$ qubits: physical for $\displaystyle \mathcal{C}_{l+1}$, logical for $\displaystyle \mathcal{C}_{l}$ };
\draw (62,190) node [anchor=north west][inner sep=0.75pt]   [align=left] {level $\displaystyle l-1$ qubits: physical for $\displaystyle \mathcal{C}_{l}$};

\end{tikzpicture}
    \caption{\label{fig:concatenation}(inspired by Fig. 9.1 of~\cite{gottesman2024surviving}) A logical qubit at level $l+1$ of the concatenation is encoded into $n_{l+1}$ physical qubits of level $l$ using the \emph{inner} code $\mathcal{C}_{l+1}$ with code parameters $((n_{l+1},2))$. These level-$l$ qubits are furthermore logical qubits of the \emph{outer} code $\mathcal{C}_{l}$ with code parameters $((n_l,2))$, which individually are encoded into $n_l$ physical qubits at level $l-1$. This construction produces a concatenated code with parameters $((n_{l+1}n_l,2))$, where qubits at level $l=0$ are the physical qubits on actual the hardware. To ensure fault-tolerance, the respective encodings are typically applied bottom-to-top, as also indicated in \cref{fig:figure1}.}
\end{figure}

\subsection{\label{subsec:concatenation}Code Concatenation}

The underlying idea of code concatenation is to improve the logical error rate by recursively encoding the state into two or even multiple quantum error correction codes~\cite{knill1996concatenated}. This is achieved by encoding the logical state using an \emph{inner} code, for which its physical qubits are the logical qubits of multiple instances of an \emph{outer} code. This procedure is sketched for \emph{inner} codes $\mathcal{C}_{l+1}$ with hyperparameters $((n_{l+1},2))$ and \emph{outer} codes $\mathcal{C}_{l}$ with $((n_l,2))$ at arbitrary concatenation levels $l$ in \cref{fig:concatenation}, where w.l.o.g.\ we restrict to code patches with only single logical qubits for the remainder of this work~\cite{yoshida2025concatenate}.

 Given encoders $\mathrm{Enc}_{l+1}$ for $\mathcal{C}_{l+1}$ and $\mathrm{Enc}_l$ for $\mathcal{C}_l$, the concatenated code is therefore produced using $\mathrm{Enc}_{l+1} \circ \mathrm{Enc}_{l}^{\otimes n_{l+1}}$, as depicted in \cref{fig:figure1}. It is easy to see that the concatenated code has code parameters
 \begin{align}
    \label{eq:concatenation_general}
     ((n_{l+1}n_l,2)),
 \end{align}
 for which it is known that $d \geq d_{l+1}d_l$, where $d_{l+1}$ and $d_l$ are the code distances of $\mathcal{C}_{l+1}$ and $\mathcal{C}_l$, respectively~\cite{gottesman2024surviving}. For stabilizer codes with parameters $[[n_{l+1},1,d_{l+1}]]$ and $[[n_l,1,d_l]]$ this simplifies to
\begin{align}
    \label{eq:concatenation_stabilizer}
    [[n_{l+1}n_l,1,d_{l+1}d_l]]
\end{align}
for the concatenated code.
It is well known that, for physical error rates below the threshold and with fully fault-tolerant gadgets, code concatenation suppresses logical error rates doubly exponentially in the concatenation level~\cite{knill1998resilient,aliferis2005quantum}. 


\section{\label{sec:method}Concatenating Variational Codes}

To scale the correction capabilities of the \gls{varqec} codes, we propose to concatenate instances that are tailored for the noise structures at every concatenation level. In particular, this requires first estimating the effective single-qubit noise channel induced by the code at the current level of the concatenation, which we discuss in \cref{subsec:analyze_concatenation}. Consecutively, we then demonstrate how to tailor the encoding-recovery pair for the next level to this noise structure in \cref{subsec:tailoring_concatenation}. 

\subsection{\label{subsec:analyze_concatenation}Analyzing Noise Channels under Concatenation}

\begin{figure}[tb]
    \centering
    \includegraphics{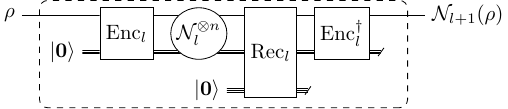}
    \caption{\label{fig:noise_tomography}Pipeline to analyze the effective channel introduced by level $l$ of a concatenated code. An input state $\rho$ from a unitary two-design undergoes encoding, noise, a single round of recovery, and decoding. Parameters $p_X, p_Y, p_Z$ of the Pauli channel in \cref{eq:pauli_noise} are (least-squares) fitted to estimate $\mathcal{N}_{l+1}$.}
\end{figure}

To analyze the noise suppression under code concatenation, we require a procedure that reconstructs the effect of a level-$l$ code on the noise strength and structure. For this, we estimate the effective single-qubit channel $\mathcal{N}_{l+1}$ by fitting a Pauli channel to fidelities computed over a unitary two-design, where each input state $\rho$ evolves governed by \cref{eq:encoding,eq:recovery,eq:decoding}. This setup is sketched in \cref{fig:noise_tomography}, and follows the standard encode-noise-recover-decode mapping used to analyze concatenated codes, based on the insight that encoding and decoding are isometric
~\cite{knill1998resilient,aliferis2005quantum}. Our approach avoids full process tomography by only recovering the diagonal part of the channel, i.e., the diagonal of its Pauli-Liouville matrix~\cite{watrous2018theory}. In this work, we are only concerned with unital noise channels, but the concept could be extended to non-unital channels like amplitude damping by incorporating Pauli-twirling techniques~\cite{geller2013efficient}.

Concretely, we fit a least-squares estimate of the single-qubit channel in the Bloch/Pauli-Liouville representation\cite{greenbaum2015introduction}, but restrict the model to a Pauli channel and drive the regression with fidelities from a unitary two-design~\cite{dankert2009exact}. To the best of our knowledge, this fidelity-only fitting technique has not been described in literature. Let $R = \mathrm{diag}(\eta_X,\eta_Y,\eta_Z)$ be the Bloch matrix of a single-qubit Pauli channel and let $r_i = [r_{X}, r_{Y}, r_{Z}]_i$ be the Bloch vector for input state $\rho_i$ from the two-design $\mathcal{S} = \left\{ \ket{\pm X}, \ket{\pm Y}, \ket{\pm Z} \right\}$. With prediction $\frac{1}{2} (1 + r_i^T R r_i) = \frac{1}{2} (1 + \eta_X r_{X}^2 + \eta_Y r_{Y}^2 + \eta_Z r_{Z}^2)_i$ and target $b_i := 2 F(\rho_i,\hat{\rho_i}) - 1$ evaluated following \cref{fig:noise_tomography}, we can set up the least-squares formulation
\begin{align}
    \label{eq:least-squares}
    \mathrm{argmin}_{\eta} \left\| A \mathcal{\eta} - b \right\|^2.
\end{align}
Hereby, $A=[r_{X}^2, r_{Y}^2, r_{Z}^2]_i \in \mathbb{R}^{\left| \mathcal{S} \right| \times 3}$ and $b=[b_i]_i \in \mathbb{R}^{\left| \mathcal{S} \right|}$. Using the six cardinal two-design states described above, the solution to \cref{eq:least-squares} is given by
\begin{align}
    \label{eq:least-squares-solution}
    \eta_{P} = F(\ket{+P},\widehat{\ket{+P}}) +  F(\ket{-P},\widehat{\ket{-P}}) - 1,
\end{align}
where $P \in \left\{ X, Y, Z \right\}$ and $F$ is the measured fidelity. The probabilities appearing in the Kraus representation from \cref{eq:pauli_noise} follow as $p_X = \frac{1}{4} (1 - \eta_Y - \eta_Z + \eta_X)$, and accordingly for $p_Y$ and $p_Z$. In case the effective channel $\mathcal{N}_{l+1}$ is exactly represented by single-qubit Pauli noise, we are done. Otherwise, the solutions following \cref{eq:least-squares-solution} might not satisfy the properties of a \gls{cptp} map, in particular violate $p_P \geq 0$ for all $P$, and $p_X + p_Y + p_Z \leq 1$. In such cases, we can recover the closest physically valid Pauli channel by mapping to the \emph{probability simplex} of $p_X, p_Y, p_Z$ with standard sorting-based approaches~\cite{duchi2008efficient}.

\subsection{\label{subsec:tailoring_concatenation}Tailoring Codes for Concatenation Level}

\begin{figure}[tb]
    \centering
    \includegraphics{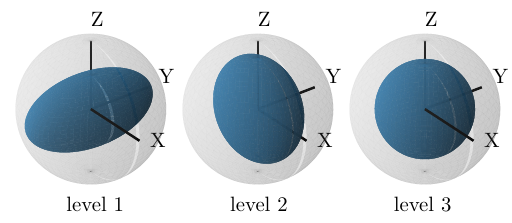}
    \caption{\label{fig:concatenate_bloch}Shift of noise structure under concatenation of (i) a $((5,2))$ \gls{varqec} code tailored to $\mathcal{N}_{\mathrm{yflip}}$ noise at level $1$, resulting in Pauli noise featuring only disjunct bit- and phaseflips at level $2$; (ii) another $((5,2))$ \gls{varqec} code tailored to this noise structure, resulting in almost uniform depolarizing noise at level $3$. For the next level, as the noise is mostly unstructured, we continue with concatenating a standard $[[5,1,3]]$ code. For noise suppression behaviour see~\cref{fig:concatenate_combined}, and for details on the noise channels see~\cref{tab:concatenation_noise_combined}.}
\end{figure}

\afterpage{
\noindent
\begin{figure*}
    \centering
    \begin{minipage}[t]{249pt}
        \centering\vspace{0pt}
        \includegraphics{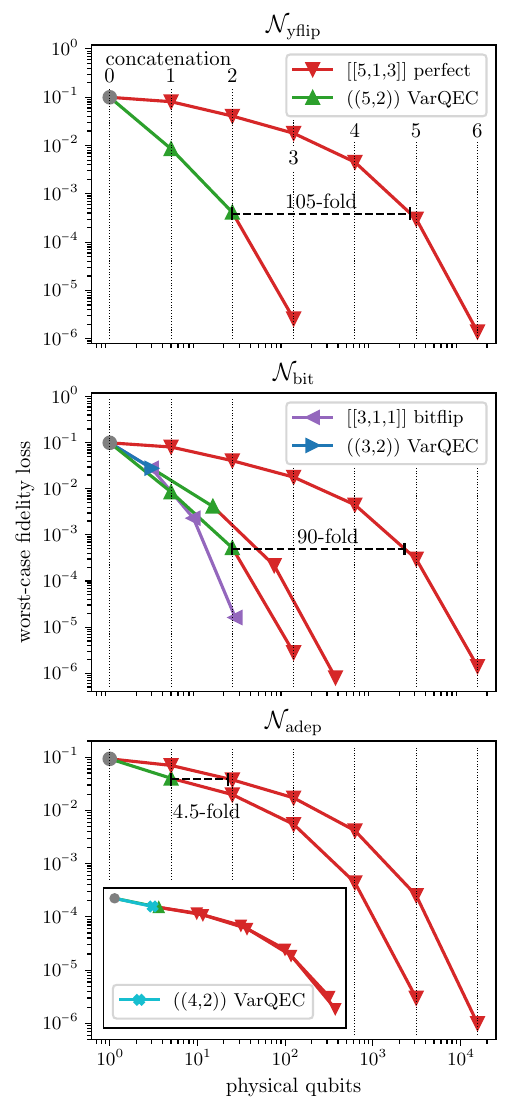}
        \captionof{figure}{\label{fig:concatenate_combined}Noise suppression under code concatenation. The colors and markers indicate which code was concatenated at which level, with \gls{varqec} codes targeted to the specific noise structures. As most channels converge to uniform depolarizing noise under concatenation (see \cref{subsec:tailoring_concatenation}), we concatenate with the standard $[[5,1,3]]$ code, once the structure is insufficient for tailoring, at which point we also show the overhead reduction compared to concatenating only standard stabilizer codes from the beginning. Details on the noise at every level of the concatenation are to be found in \cref{tab:concatenation_noise_combined}.}
    \end{minipage}
    \hspace{10pt}
    \begin{minipage}[t]{249pt}
        \centering\vspace{0pt}
        {\footnotesize
            \begin{tabular}{cc|cccc}
    \toprule
    \multirow{2}{*}{\textbf{code}} & \multirow{2}{*}{\textbf{level}} & \textbf{strength} & \multicolumn{3}{c}{\textbf{proportion}} \\
    & & $p$ & $\nicefrac{p_X}{p}$ & $\nicefrac{p_Y}{p}$ & $\nicefrac{p_Z}{p}$ \\
    \midrule
    $\mathcal{N}_{\mathrm{yflip}}$ & 0 & $0.10$ & $0.00$ & $1.00$ & $0.00$ \\
    \midrule
    \multirow{6}{*}{$[[5,1,3]]$} & 1 & $0.082$ & $0.50$ & $0.00$ & $0.50$ \\
    & 2 & $0.056$ & $0.37$ & $0.24$ & $0.37$ \\
    & 3 & $0.027$ & $\nicefrac{1}{3}$ & $\nicefrac{1}{3}$ & $\nicefrac{1}{3}$ \\
    & 4 & $0.0068$ & $\nicefrac{1}{3}$ & $\nicefrac{1}{3}$ & $\nicefrac{1}{3}$ \\
    & 5 & $0.00046$ & $\nicefrac{1}{3}$ & $\nicefrac{1}{3}$ & $\nicefrac{1}{3}$ \\
    & 6 & $0.0000020$ & $\nicefrac{1}{3}$ & $\nicefrac{1}{3}$ & $\nicefrac{1}{3}$ \\
    \midrule
    \multirow{1}{*}{$((5,2))$} & 1 & $0.0085$ & $0.50$ & $0.00$ & $0.50$ \\
    \multirow{1}{*}{$((5,2))$} & 2 & $0.00060$ & $0.33$ & $0.33$ & $0.34$ \\
    \multirow{1}{*}{$[[5,1,3]]$} & 3 & $0.0000037$ & $\nicefrac{1}{3}$ & $\nicefrac{1}{3}$ & $\nicefrac{1}{3}$ \\
    \bottomrule
    
    \multicolumn{6}{c}{}\\
    
    \toprule
    \multirow{2}{*}{\textbf{code}} & \multirow{2}{*}{\textbf{level}} & \textbf{strength} & \multicolumn{3}{c}{\textbf{proportion}} \\
    & & $p$ & $\nicefrac{p_X}{p}$ & $\nicefrac{p_Y}{p}$ & $\nicefrac{p_Z}{p}$ \\
    \midrule
    $\mathcal{N}_{\mathrm{bit}}$ & 0 & $0.10$ & $1.00$ & $0.00$ & $0.00$ \\
    \midrule
    \multirow{4}{*}{$[[3,1,1]]$} & 1 & $0.028$ & $1.00$ & $0.00$ & $0.00$ \\
    & 2 & $0.0023$ & $1.00$ & $0.00$ & $0.00$ \\
    & 3 & $0.000016$ & $1.00$ & $0.00$ & $0.00$ \\
    & 4 & $0.0000001$ & $1.00$ & $0.00$ & $0.00$ \\
    \midrule
    \multirow{6}{*}{$[[5,1,3]]$} & 1 & $0.082$ & $0.00$ & $0.50$ & $0.50$ \\
    & 2 & $0.055$ & $0.26$ & $0.37$ & $0.37$ \\
    & 3 & $0.027$ & $\nicefrac{1}{3}$ & $\nicefrac{1}{3}$ & $\nicefrac{1}{3}$ \\
    & 4 & $0.0068$ & $\nicefrac{1}{3}$ & $\nicefrac{1}{3}$ & $\nicefrac{1}{3}$ \\
    & 5 & $0.00045$ & $\nicefrac{1}{3}$ & $\nicefrac{1}{3}$ & $\nicefrac{1}{3}$ \\
    & 6 & $0.0000020$ & $\nicefrac{1}{3}$ & $\nicefrac{1}{3}$ & $\nicefrac{1}{3}$ \\
    \midrule
    \multirow{1}{*}{$((3,2))$} & 1 & $0.028$ & $0.50$ & $0.50$ & $0.00$ \\
    \multirow{1}{*}{$((5,2))$} & 2 & $0.0057$ & $0.27$ & $0.30$ & $0.43$ \\
    \multirow{1}{*}{$[[5,1,3]]$} & 3 & $0.00033$ & $\nicefrac{1}{3}$ & $\nicefrac{1}{3}$ & $\nicefrac{1}{3}$ \\
    \multirow{1}{*}{$[[5,1,3]]$} & 4 & $0.000011$ & $\nicefrac{1}{3}$ & $\nicefrac{1}{3}$ & $\nicefrac{1}{3}$ \\
    \midrule
    \multirow{1}{*}{$((5,2))$} & 1 & $0.0086$ & $0.00$ & $0.50$ & $0.50$ \\
    \multirow{1}{*}{$((5,2))$} & 2 & $0.00065$ & $0.20$ & $0.35$ & $0.45$ \\
    \multirow{1}{*}{$[[5,1,3]]$} & 3 & $0.0000043$ & $\nicefrac{1}{3}$ & $\nicefrac{1}{3}$ & $\nicefrac{1}{3}$ \\
    \bottomrule

    \multicolumn{6}{c}{}\\

    \toprule
    \multirow{2}{*}{\textbf{code}} & \multirow{2}{*}{\textbf{level}} & \textbf{strength} & \multicolumn{3}{c}{\textbf{proportion}} \\
    & & $p$ & $\nicefrac{p_X}{p}$ & $\nicefrac{p_Y}{p}$ & $\nicefrac{p_Z}{p}$ \\
    \midrule
    $\mathcal{N}_{\mathrm{adep}}$ & 0 & $0.10$ & $0.07$ & $0.07$ & $0.86$ \\
    \midrule
    \multirow{6}{*}{$[[5,1,3]]$} & 1 & $0.081$ & $0.44$ & $0.44$ & $0.12$ \\
    & 2 & $0.054$ & $0.35$ & $0.35$ & $0.30$ \\
    & 3 & $0.026$ & $\nicefrac{1}{3}$ & $\nicefrac{1}{3}$ & $\nicefrac{1}{3}$ \\
    & 4 & $0.0062$ & $\nicefrac{1}{3}$ & $\nicefrac{1}{3}$ & $\nicefrac{1}{3}$ \\
    & 5 & $0.00038$ & $\nicefrac{1}{3}$ & $\nicefrac{1}{3}$ & $\nicefrac{1}{3}$ \\
    & 6 & $0.0000014$ & $\nicefrac{1}{3}$ & $\nicefrac{1}{3}$ & $\nicefrac{1}{3}$ \\
    \midrule
    \multirow{1}{*}{$((4,2))$} & 1 & $0.062$ & $0.36$ & $0.32$ & $0.32$ \\
    \multirow{1}{*}{$[[5,1,3]]$} & 2 & $0.035$ & $\nicefrac{1}{3}$ & $\nicefrac{1}{3}$ & $\nicefrac{1}{3}$ \\
    \multirow{1}{*}{$[[5,1,3]]$} & 3 & $0.011$ & $\nicefrac{1}{3}$ & $\nicefrac{1}{3}$ & $\nicefrac{1}{3}$ \\
    \multirow{1}{*}{$[[5,1,3]]$} & 4 & $0.0012$ & $\nicefrac{1}{3}$ & $\nicefrac{1}{3}$ & $\nicefrac{1}{3}$ \\
    \multirow{1}{*}{$[[5,1,3]]$} & 5 & $0.000014$ & $\nicefrac{1}{3}$ & $\nicefrac{1}{3}$ & $\nicefrac{1}{3}$ \\
    \midrule
    \multirow{1}{*}{$((5,2))$} & 1 & $0.057$ & $0.34$ & $0.34$ & $0.32$ \\
    \multirow{1}{*}{$[[5,1,3]]$} & 2 & $0.030$ & $\nicefrac{1}{3}$ & $\nicefrac{1}{3}$ & $\nicefrac{1}{3}$ \\
    \multirow{1}{*}{$[[5,1,3]]$} & 3 & $0.0081$ & $\nicefrac{1}{3}$ & $\nicefrac{1}{3}$ & $\nicefrac{1}{3}$ \\
    \multirow{1}{*}{$[[5,1,3]]$} & 4 & $0.00066$ & $\nicefrac{1}{3}$ & $\nicefrac{1}{3}$ & $\nicefrac{1}{3}$ \\
    \multirow{1}{*}{$[[5,1,3]]$} & 5 & $0.0000044$ & $\nicefrac{1}{3}$ & $\nicefrac{1}{3}$ & $\nicefrac{1}{3}$ \\
    \bottomrule
\end{tabular}

        }
        \renewcommand{\tablename}{Tab.}
        \captionof{table}{\label{tab:concatenation_noise_combined}Change of noise structure under code concatenation. To unify the notation, we note the total noise strength $p$ and the $X$, $Y$, and $Z$ contributions of single-qubit Pauli noise, as defined in \cref{eq:pauli_noise}. We list the noise strength and structure change after each concatenation level with the \gls{qec} codes specified in the leftmost column. This structure change is visualized for initially $\mathcal{N}_{\mathrm{yflip}}$ noise in \cref{fig:concatenate_bloch}, and the noise suppression under concatenation is shown in \cref{fig:concatenate_combined}.}
    \end{minipage}
\end{figure*}
}

For the first level of the concatenation, we straightforwardly use the \gls{varqec} approach described in \cref{subsec:varqec} to tailor a code $\mathcal{C}_1$, consisting of encoding $\mathrm{Enc}_1$ and recovery $\mathrm{Rec}_1$, to the initial noise channel $\mathcal{N}_1$. Afterwards, the tomography method from \cref{subsec:analyze_concatenation} is used to estimate the effective noise channel after this code instance, and therefore the \emph{input} for the next level of the concatenation, denoted as $\mathcal{N}_2$. In the next cycle, a code $\mathcal{C}_2$ is tailored to this noise, resulting in another effective channel $\mathcal{N}_3$ after encoding and correction. This procedure is iteratively repeated by tailoring a code $\mathcal{C}_l$ to noise $\mathcal{N}_l$, producing the effective channel $\mathcal{N}_{l+1}$, until a desired target noise suppression rate is achieved.

It is crucial to tailor the codes to every level of the concatenation separately, as not only the noise strength, but also the noise structure changes after each concatenation instance~\cite{huang2019robustness}. Using different codes for each concatenation level has been proven effective in reducing the overall qubit count, even when restricting to stabilizer codes~\cite{yoshida2025concatenate}.
In particular, it has been observed that under non-CSS codes, anisotropic noise quickly changes to effectively isotropic channels under concatenation, i.e., converges towards symmetric depolarizing noise as given in \cref{eq:noise_depolarizing}~\cite{huang2019robustness}. All codes we consider in this work are non-CSS, including the $[[3,1,1]]$ \emph{bitflip} and $[[5,1,3]]$ \emph{perfect} stabilizer codes~\cite{shor1995scheme,laflamme1996perfect}, as well as all \gls{varqec} codes, as they are non-additive. This behavior is expected to be different for CSS codes like e.g.\ the $[[7,1,3]]$ \emph{Steane} or the $[[9,1,3]]$ \emph{Shor} code~\cite{steane1996error,shor1995scheme}, where anisotropic noise typically remains anisotropic under concatenation~\cite{huang2019robustness}. However, for the sake of this paper, the $[[5,1,3]]$ code poses the hardest baseline, as it corrects for an arbitrary single-qubit error with the provably smallest number of physical qubits, which keeps the resource overhead analyzed in \cref{sec:empirical} minimal. To ensure fault tolerance in practice, one needs to encode using logical operations within the codespace of the previous concatenation level, which typically leads to still choosing the $[[7,1,3]]$ code over $[[5,1,3]]$. However, such considerations, also for the \gls{varqec} codes, are out of the scope of this manuscript and will be investigated in future work~\cite{meyer2026learningb}.


\section{\label{sec:empirical}Empirical Setup and Evaluation}

To empirically analyze the behaviour of a noise channel under code concatenation, we implement the noise tomography procedure described in \cref{subsec:analyze_concatenation} in \texttt{python}, relying mainly on the \texttt{pennylane}~\cite{bergholm2018pennylane} and \texttt{qiskit-torch-module}~\cite{meyer2024qiskit} libraries. For tailoring the \gls{varqec} codes to the specific noise structures, we employed an open-source implementation of the \gls{varqec} approach~\cite{meyer2025learning}. Training is conducted using the quasi-Newton L-BFGS optimizer~\cite{liu1989limited} on \gls{rea} ans{\"a}tze~\cite{meyer2025learning} with $n \cdot (n+1)$ blocks for the encoding unitary. While training \gls{varqec} codes for every concatenation level from scratch is possible, we observed that using warm-start initialization~\cite{meyer2024warm} with parameters from the previous level speeds up the convergence significantly. For our experiments, we simulate the encoding and recovery operations themselves to be noise-free, i.e.\ we assume the existence of fault-tolerant gadgets for realizing encoding and correcting.

Using this setup, in \cref{fig:concatenate_combined} we show the analysis of three noise channels with initial noise strength $p=0.1$. For each, we compare fully stabilizer-code concatenations with hybrid concatenations that use tailored \gls{varqec} codes at the outer and stabilizer codes at the inner levels. \cref{fig:concatenate_combined} shows the resulting suppression by plotting the worst-case fidelity loss (\cref{eq:floss_worst}) versus the number of physical qubits $n$ following \cref{eq:concatenation_general,eq:concatenation_stabilizer}. Noise structures at each level are listed in \cref{tab:concatenation_noise_combined}.

For initial Pauli $Y$-flip $\mathcal{N}_{\mathrm{yflip}}$ noise (top plot of \cref{fig:concatenate_combined}), we compare concatenating $((5,2))$ \gls{varqec} codes at level $1$ and $2$ and a standard $[[5,1,3]]$ code at level $3$ to using the $[[5,1,3]]$ codes at every level. In the former case, we switch to the standard code at level $3$ because the effective noise is nearly uniformly depolarizing with $\frac{p_X}{p} \approx \frac{p_Y}{p} \approx \frac{p_Z}{p} \approx \frac{1}{3}$ (see also \cref{fig:concatenate_bloch}), leaving no exploitable structure for the \gls{varqec} procedure. For comparable noise suppression, the hybrid strategy reduces resources by about $105$-fold, from $n \approx 2625$ to $n=25$ (i.e., $5^2$ after two levels) physical qubits.

For initial bitflip $\mathcal{N}_{\mathrm{bit}}$ noise (middle plot of \cref{fig:concatenate_combined}), concatenating tailored $5$-qubit codes yield a substantial $90$-fold overhead reduction. We also test $3$-qubit codes at level $1$: the $[[3,1,1]]$ bitflip code and $((3,2))$ \gls{varqec} codes. Both achieve the same total noise suppression after the first layer, but the variational code alters the noise structure to Pauli noise with $\frac{p_X}{p} \approx \frac{p_Y}{p} \approx \frac{1}{2}$, whereas the stabilizer code maintains an effective single-qubit bitflip channel. This, however, requires concatenating the $((3,2))$ with a higher-qubit $((5,2))$ code at level $2$, to achieve beyond break-even noise suppression, increasing overall resource requirements. Thus, stabilizer codes tailored to noise structure can still outperform machine-learned non-additive codes in the concatenated setup. However, this also points to a natural future extension of the concept: instead of minimizing the worst-case fidelity loss, the objective could be to enforce desired channel structure after correction, enabling not only suppression but finer-grained noise control.

Lastly, for asymmetric depolarizing noise $\mathcal{N}_{\mathrm{adep}}$ (bottom plot of \cref{fig:concatenate_combined}), a single instance of a tailored $((5,2))$ \gls{varqec} code already makes the effective noise channel almost uniform. Therefore, from the second level onwards, we concatenate with standard stabilizer codes. The resulting overhead reduction is smaller, but still significant at $4.5$-fold. Using a $((4,2))$ \gls{varqec} code at the initial level yields a similar picture (see inset).

Overall, across considered noise structures, combining variational and standard codes reduces overhead by one to two orders of magnitude compared with stabilizer-only concatenations. In \cref{fig:concatenate_combined}, both axes are logarithmic, while concatenation levels are shown on a linear scale. In all cases, we observe the theoretically predicted asymptotically super-exponential suppression of noise under concatenation~\cite{knill1998resilient,aliferis2005quantum}, with the discussed significant practical scaling differences.


\glsresetall
\section{\label{sec:discussion}Discussion and Outlook}

This work presents a pipeline that concatenates non-additive and additive codes, tailoring each level to the effective noise structure. To enable this, we introduced an efficient fidelity-only, two-design estimator that recovers the Pauli-diagonal part of the encode-noise-recover-decode channel. Based upon this, a learning-based procedure is employed to construct \gls{varqec} codes~\cite{meyer2025learning} for the concatenation levels where noise is sufficiently structured, and we resorted to standard stabilizer codes in regimes of near-uniform noise. Across different initial noise channels, layer-wise tailoring yields overhead reductions compared to concatenating just stabilizer codes by one to two orders of magnitude, demonstrating the theoretically predicted double-exponentially noise suppression.

We restrict our analysis to Pauli noise channels, while noting that the underlying noise-tomography procedure can be extended, for example via Pauli twirling, to capture non-unital channels such as amplitude damping~\cite{meyer2025learning}. Such extensions would enable a more realistic characterization of noise reshaping under concatenation. We also focus on codes encoding a single logical qubit per patch, although the framework is conceptually not limited to this setting and can be generalized to higher-rate codes~\cite{yoshida2025concatenate}. A central limitation of the present manuscript is that it does not yet address the early fault-tolerant realization of logical operations, which must be implemented within the codespace of lower concatenation levels. This gap has conceptually been addressed in follow-up work through a co-design approach that jointly optimizes encoding procedures and ensures the existence of low-depth logical operations~\cite{meyer2026learningb}, while a realistic analysis of these operations and their implications for practical concatenation and thresholds remains for future work.

Another further promising future research direction is the modification of the loss function to enforce desired structural properties of the effective noise channel, as currently the advantage of the \gls{varqec} codes is limited due to fast convergence towards isotropic noise under concatenation.

In conclusion, concatenating level-wise noise-tailored codes allows for substantial overhead reduction under structured noise. Extending to higher-rate codes and employing the support of fault-tolerant gadgets can evolve the concept to a practical alternative for early fault-tolerant quantum computing.

\section*{Acknowledgment}
The authors used GPT-5.1 for language editing of the paper. All content was reviewed and edited by the authors, who take full responsibility for the final work.

\section*{Data Availability}
The error-correcting codes tailored to the noise structures under consideration and their concatenation levels, as well as the analysis protocol for estimating the noise structure, are available at \href{https://github.com/nicomeyer96/learning-to-concatenate}{https://github.com/nicomeyer96/learning-to-concatenate}. Additional information and data are available upon reasonable request.

\newpage
\bibliographystyle{IEEEtran}
\bibliography{paper}

@article{meyer2025learning,
  title={Learning {E}ncodings by {M}aximizing {S}tate {D}istinguishability: {V}ariational {Q}uantum {E}rror {C}orrection},
  author={Meyer, Nico and Mutschler, Christopher and Maier, Andreas and Scherer, Daniel D},
  journal={arXiv:2506.11552},
  volume={},
  number={},
  pages={},
  year={2025},
  doi={10.48550/arXiv.2506.11552}
}

@inproceedings{meyer2025variational,
  title={Variational {Q}uantum {E}rror {C}orrection},
  author={Meyer, Nico and Mutschler, Christopher and Maier, Andreas and Scherer, Daniel D},
  booktitle={IEEE International Conference on Quantum Computing and Engineering (QCE)},
  volume={2},
  number={},
  pages={456--457},
  year={2025},
  doi={10.1109/QCE65121.2025.10393}
}

@article{meyer2026learningb,
  title={Learning {L}ogical {O}perations for {A}rbitrary {Q}uantum {E}rror {C}orrection {C}odes},
  author={Meyer, Nico and Mutschler, Christopher and Seu{\ss}, Dominik and Maier, Andreas and Scherer, Daniel D},
  journal={arXiv:2605.28162},
  volume={},
  number={},
  pages={},
  year={2026},
  doi={10.48550/arXiv.2605.28162}
}

@article{fosel2018reinforcement,
  title={Reinforcement {L}earning with {N}eural {N}etworks for {Q}uantum {F}eedback},
  author={F{\"o}sel, Thomas and Tighineanu, Petru and Weiss, Talitha and Marquardt, Florian},
  journal={Phys. Rev. X},
  volume={8},
  number={3},
  pages={031084},
  year={2018},
  doi={10.1103/PhysRevX.8.031084}
}

@article{johnson2017qvector,
  title={{QVECTOR}: an algorithm for device-tailored quantum error correction},
  author={Johnson, Peter D and Romero, Jonathan and Olson, Jonathan and Cao, Yudong and Aspuru-Guzik, Al{\'a}n},
  journal={arXiv:1711.02249},
  volume={},
  number={},
  pages={},
  year={2017},
  doi={10.48550/arXiv.1711.02249}
}

@article{olle2024simultaneous,
  title={Simultaneous discovery of quantum error correction codes and encoders with a noise-aware reinforcement learning agent},
  author={Oll{\'e}, Jan and Zen, Remmy and Puviani, Matteo and Marquardt, Florian},
  journal={npj Quantum Inf.},
  volume={10},
  number={1},
  pages={1--17},
  year={2024},
  doi={10.1038/s41534-024-00920-y}
}

@article{cerezo2021variational,
  title={Variational quantum algorithms},
  author={Cerezo, Marco and Arrasmith, Andrew and Babbush, Ryan and Benjamin, Simon C and Endo, Suguru and Fujii, Keisuke and McClean, Jarrod R and Mitarai, Kosuke and Yuan, Xiao and Cincio, Lukasz and others},
  journal={Nature Rev. Phys.},
  volume={3},
  number={9},
  pages={625--644},
  year={2021},
  doi={10.1038/s42254-021-00348-9}
}

@article{liu1989limited,
  title={On the limited memory {BFGS} method for large scale optimization},
  author={Liu, Dong C and Nocedal, Jorge},
  journal={Math. Program.},
  volume={45},
  number={1},
  pages={503--528},
  year={1989},
  doi={10.1007/BF01589116}
}

@article{heussen2024measurement,
  title={Measurement-{F}ree {F}ault-{T}olerant {Q}uantum {E}rror {C}orrection in {N}ear-{T}erm {D}evices},
  author={Heu{\ss}en, Sascha and Locher, David F and M{\"u}ller, Markus},
  journal={PRX Quantum},
  volume={5},
  number={1},
  pages={010333},
  year={2024},
  doi={10.1103/PRXQuantum.5.010333}
}

@article{schumacher2002approximate,
  title={Approximate {Q}uantum {E}rror {C}orrection},
  author={Schumacher, Benjamin and Westmoreland, Michael D},
  journal={Quantum Inf. Process.},
  volume={1},
  number={},
  pages={5--12},
  year={2002},
  doi={10.1023/A:1019653202562}
}

@article{laflamme1996perfect,
  title={Perfect {Q}uantum {E}rror {C}orrecting {C}ode},
  author={Laflamme, Raymond and Miquel, Cesar and Paz, Juan Pablo and Zurek, Wojciech Hubert},
  journal={Phys. Rev. Lett.},
  volume={77},
  number={1},
  pages={198},
  year={1996},
  doi={10.1103/PhysRevLett.77.198}
}

@inproceedings{meyer2024qiskit,
  title={Qiskit-{T}orch-{M}odule: {F}ast {P}rototyping of {Q}uantum {N}eural {N}etworks},
  author={Meyer, Nico and Ufrecht, Christian and Periyasamy, Maniraman and Plinge, Axel and Mutschler, Christopher and Scherer, Daniel D and Maier, Andreas},
  booktitle={IEEE International Conference on Quantum Computing and Engineering (QCE)},
  volume={1},
  number={},
  pages={817--823},
  year={2024},
  doi={10.1109/QCE60285.2024.00101}
}

@article{bergholm2018pennylane,
  title={Penny{L}ane: {A}utomatic differentiation of hybrid quantum-classical computations},
  author={Bergholm, Ville and Izaac, Josh and Schuld, Maria and Gogolin, Christian and Ahmed, Shahnawaz and Ajith, Vishnu and Alam, M Sohaib and Alonso-Linaje, Guillermo and AkashNarayanan, B and Asadi, Ali and others},
  journal={arXiv:1811.04968},
  volume={},
  number={},
  year={2018},
  doi={10.48550/arXiv.1811.04968}
}

@article{kraus1971general,
  title={General state changes in quantum theory},
  author={Kraus, Karl},
  journal={Ann. Phys.},
  volume={64},
  number={2},
  pages={311--335},
  year={1971},
  doi={10.1016/0003-4916(71)90108-4}
}

@article{shor1995scheme,
  title={Scheme for reducing decoherence in quantum computer memory},
  author={Shor, Peter W},
  journal={Phys. Rev. A},
  volume={52},
  number={4},
  pages={R2493},
  year={1995},
  doi={10.1103/PhysRevA.52.R2493}
}

@article{steane1996error,
  title={Error {C}orrecting {C}odes in {Q}uantum {T}heory},
  author={Steane, Andrew M},
  journal={Phys. Rev. Lett.},
  volume={77},
  number={5},
  pages={793},
  year={1996},
  doi={10.1103/PhysRevLett.77.793}
}

@book{hastie2009elements,
  title={The {E}lements of {S}tatistical {L}earning: {D}ata {M}ining, {I}nference, and {P}rediction},
  author={Hastie, Trevor},
  year={2009},
  publisher={Springer},
  doi={10.1007/978-0-387-84858-7}
}

@inproceedings{meyer2024warm,
  title={Warm-{S}tart {V}ariational {Q}uantum {P}olicy {I}teration},
  author={Meyer, Nico and Murauer, Jakob and Popov, Alexander and Ufrecht, Christian and Plinge, Axel and Mutschler, Christopher and Scherer, Daniel D},
  booktitle={IEEE International Conference on Quantum Computing and Engineering (QCE)},
  volume={1},
  number={},
  pages={1458--1466},
  year={2024},
  doi={10.1109/QCE60285.2024.00172}
}

@article{olle2025scaling,
  title={Scaling the {A}utomated {D}iscovery of {Q}uantum {C}ircuits via {R}einforcement {L}earning with {G}adgets},
  author={Oll{\'e}, Jan and Yevtushenko, Oleg M and Marquardt, Florian},
  journal={arXiv:2503.11638},
  volume={},
  number={},
  pages={},
  year={2025},
  doi={10.48550/arXiv.2503.1163}
}

@article{nautrup2019optimizing,
  title={Optimizing {Q}uantum {E}rror {C}orrection {C}odes with {R}einforcement {L}earning},
  author={Nautrup, Hendrik Poulsen and Delfosse, Nicolas and Dunjko, Vedran and Briegel, Hans J and Friis, Nicolai},
  journal={Quantum},
  volume={3},
  number={},
  pages={215},
  year={2019},
  doi={10.22331/q-2019-12-16-215}
}

@article{locher2023quantum,
  title={Quantum {E}rror {C}orrection with {Q}uantum {A}utoencoders},
  author={Locher, David F and Cardarelli, Lorenzo and M{\"u}ller, Markus},
  journal={Quantum},
  volume={7},
  number={},
  pages={942},
  year={2023},
  doi={10.22331/q-2023-03-09-942}
}

@article{katabarwa2024early,
  title={Early {F}ault-{T}olerant {Q}uantum {C}omputing},
  author={Katabarwa, Amara and Gratsea, Katerina and Caesura, Athena and Johnson, Peter D},
  journal={PRX Quantum},
  volume={5},
  number={2},
  pages={020101},
  year={2024},
  doi={10.1103/PRXQuantum.5.020101}
}

@article{knill1996concatenated,
  title={Concatenated {Q}uantum {C}odes},
  author={Knill, Emanuel and Laflamme, Raymond},
  journal={arXiv:quant-ph/9608012},
  volume={},
  number={},
  pages={},
  year={1996},
  doi={10.48550/arXiv.quant-ph/9608012}
}

@misc{gottesman2024surviving,
  title={Surviving as a {Q}uantum {C}omputer in a {C}lassical {W}orld},
  author={Gottesman, Daniel},
  year={2024},
  url={https://www.cs.umd.edu/class/spring2024/cmsc858G/QECCbook-2024-ch1-15.pdf},
  note={lecture notes}
}

@article{knill1998resilient,
  title={Resilient {Q}uantum {C}omputation},
  author={Knill, Emanuel and Laflamme, Raymond and Zurek, Wojciech H},
  journal={Science},
  volume={279},
  number={5349},
  pages={342--345},
  year={1998},
  doi={10.1126/science.279.5349.342}
}

@article{aliferis2005quantum,
  title={Quantum accuracy threshold for concatenated distance-3 codes},
  author={Aliferis, Panos and Gottesman, Daniel and Preskill, John},
  journal={Quantum Inf. Comput.},
  volume={6},
  number={2},
  pages={97-165},
  year={2005},
  doi={10.5555/2011665.2011666}
}

@article{yoshida2025concatenate,
  title={Concatenate codes, save qubits},
  author={Yoshida, Satoshi and Tamiya, Shiro and Yamasaki, Hayata},
  journal={npj Quantum Inf.},
  volume={11},
  number={1},
  pages={88},
  year={2025},
  doi={10.1038/s41534-025-01035-8}
}

@inproceedings{ambainis2007quantum,
  title={Quantum t-designs: t-wise {I}ndependence in the {Q}uantum {W}orld},
  author={Ambainis, Andris and Emerson, Joseph},
  booktitle={IEEE Conference on Computational Complexity (CCC'07)},
  volume={1},
  number={},
  pages={129--140},
  year={2007},
  doi={10.1109/CCC.2007.26}
}

@article{dankert2009exact,
  title={Exact and approximate unitary 2-designs and their application to fidelity estimation},
  author={Dankert, Christoph and Cleve, Richard and Emerson, Joseph and Livine, Etera},
  journal={Phys. Rev. A},
  volume={80},
  number={1},
  pages={012304},
  year={2009},
  doi={10.1103/PhysRevA.80.012304}
}

@book{watrous2018theory,
  title={The {T}heory of {Q}uantum {I}nformation},
  author={Watrous, John},
  year={2018},
  publisher={{Cambridge University Press}},
  doi={10.1017/9781316848142}
}

@article{greenbaum2015introduction,
  title={Introduction to {Q}uantum {G}ate {S}et {T}omography},
  author={Greenbaum, Daniel},
  journal={arXiv:1509.02921},
  volume={},
  number={},
  pages={},
  year={2015},
  doi={10.48550/arXiv.1509.02921}
}

@inproceedings{duchi2008efficient,
  title={Efficient projections onto the l 1-ball for learning in high dimensions},
  author={Duchi, John and Shalev-Shwartz, Shai and Singer, Yoram and Chandra, Tushar},
  booktitle={International Conference on Machine Learning (ICML)},
  volume={},
  number={},
  pages={272--279},
  year={2008},
  doi={10.1145/1390156.1390191}
}

@article{huang2019robustness,
  title={Robustness of the concatenated quantum error-correction protocol against noise for channels affected by fluctuation},
  author={Huang, Long and Wu, Xiaohua and Zhou, Tao},
  journal={Phys. Rev. A},
  volume={100},
  number={},
  pages={042321},
  year={2019},
  doi={10.1103/PhysRevA.100.042321}
}

@article{krantz2019quantum,
  title={A quantum engineer's guide to superconducting qubits},
  author={Krantz, Philip and Kjaergaard, Morten and Yan, Fei and Orlando, Terry P and Gustavsson, Simon and Oliver, William D},
  journal={Appl. Phys. Rev.},
  volume={6},
  number={2},
  pages={021318},
  year={2019},
  doi={10.1063/1.5089550}
}

@article{erhard2019characterizing,
  title={Characterizing large-scale quantum computers via cycle benchmarking},
  author={Erhard, Alexander and Wallman, Joel J and Postler, Lukas and Meth, Michael and Stricker, Roman and Martinez, Esteban A and Schindler, Philipp and Monz, Thomas and Emerson, Joseph and Blatt, Rainer},
  journal={Nature Commun.},
  volume={10},
  number={1},
  pages={5347},
  year={2019},
  doi={10.1038/s41467-019-13068-7}
}

@phdthesis{gottesman1997stabilizer,
  title={Stabilizer {C}odes and {Q}uantum {E}rror {C}orrection},
  author={Gottesman, Daniel},
  year={1997},
  school={{California Institute of Technology}}
}

@article{geller2013efficient,
  title={Efficient error models for fault-tolerant architectures and the Pauli twirling approximation},
  author={Geller, Michael R and Zhou, Zhongyuan},
  journal={Phys. Rev. A},
  volume={88},
  number={},
  pages={012314},
  year={2013},
  doi={10.1103/PhysRevA.88.012314}
}

@article{cao2022quantum,
  title={Quantum variational learning for quantum error-correcting codes},
  author={Cao, Chenfeng and Zhang, Chao and Wu, Zipeng and Grassl, Markus and Zeng, Bei},
  journal={Quantum},
  volume={6},
  pages={828},
  year={2022},
  doi={10.22331/q-2022-10-06-828}
}

\end{document}